\documentclass[12pt]{iopart}

\usepackage{graphicx}
\usepackage{setstack}
\usepackage{iopams}
\usepackage{amsthm}

\begin{document}

\title[SUSY,ES,SI, and Infeld-Hull Venn Diagram]{A Venn diagram for supersymmetric, exactly solvable, shape invariant, and Infeld-Hull factorizable potentials}

\author{M.Mustafa$^1$ and S.Kais$^2$}
\address{1 Department of Physics, Purdue University}
\address{2 Department of Chemistry, Purdue University}

\ead{\mailto{mustafa@purdue.edu}}

\begin{abstract}
Supersymmetry, shape invariance, exact solubility, and the factorization method are often studied together in the literature. At the dawn of these topics confusion was present in regards to their scope of applicability and the relation among them. Considerable work have been put to study and resolve the relation among two or more of these topics. These works are scattered over the literature. While looking at the literature, one can not overlook the number of places where authors confuse these terms, and concluding implications depending on wrong assumptions of the relation between two or more of these topics. In this letter we define supersymmetry, and shape invariance, and show the relations which connects them to exact solubility and the factorization method, referring to the literature for the respective detailed work and proofs. At last we conclude our letter with a Venn diagram which illustrates those relations.
\end{abstract}

\section{Supersymmetric Quantum Mechanics}\label{SUSYsection}
Witten, in his seminal paper \cite{Witten1981513}, defined the \emph{algebra} of
a supersymmetric quantum system,
which he derived from the algebra of supersymmetry in field theory.
In supersymmetric quantum systems, there are \emph{charge operators} $Q_i$ which
commute with the Hamiltonian

\begin{equation}
[Q_i,\mathcal{H}]=0,\ \ \ \ \ \ \ \ \ \ \ \ i=1,2,...,N
\end{equation}

and they obey the algebra

\begin{equation}
\{Q_i,Q_j\}=\delta_{ij}\mathcal{H}
\end{equation}

where $\mathcal{H}$ is called the \emph{Supersymmetric Hamiltonian}. Witten
stated that the simplest quantum mechanical system has $N=2$,
it was later shown that the case where $N=1$, if it is supersymmetric, it is
equivalent to an $N=2$ supersymmetric quantum system \cite{Combescur2004}.
In the case where $N=2$ we can define

\begin{equation}
Q=(\frac{1}{\sqrt{2}})(Q_1 + i Q_2), \ \ \ \ \ \ \
Q^{\dagger}=(\frac{1}{\sqrt{2}})(Q_1 - i Q_2)
\end{equation}

these operators will obey the algebra

\begin{equation}
\label{algebra}
 \mathcal{H}=\{Q,Q^{\dagger}\},\ \ \ \ \ \ Q^2=Q^{\dagger2}=0
\end{equation}
 
consequently,

\begin{equation}
\label{qhcommutator}
 [Q,\mathcal{H}]=[Q^{\dagger},\mathcal{H}]=0
\end{equation}

We can represent our charge operators as $2\times2$ matrices

\numparts
 \label{charges}
\begin{equation}
 Q=\left(
\begin{array}{cc}
0 & 0 \\
A & 0
\end{array} \right)
\end{equation}

 \begin{equation} Q^{\dagger}=\left(
\begin{array}{cc}
0 & A^{\dagger} \\
0 & 0
\end{array} \right)
\end{equation}
\endnumparts

where $A$ is a linear operator, and $A^{\dagger}$ is its adjoint. Now the
supersymmetric Hamiltonian has the representation

\begin{equation}
 \mathcal{H}=\{Q,Q^{\dagger}\}=\left(
\begin{array}{cc}
 A^{\dagger}A & 0\\
0 & AA^{\dagger}
\end{array}
\right)=\left(
\begin{array}{cc}
 H_{-} & 0\\
0 & H_{+}
\end{array}
\right)
\end{equation}

where in the last step we designated the hermitian operators $A^{\dagger}A$ and
$AA^{\dagger}$ as the Hamiltonians $H_{-}$ and $H_{+}$, respectively.
It is important to notice that $H_{-}$ and $H_{+}$ are positive semi-definite. for if $|\alpha\left.\right>$ is an eigenstate in the Hilbert space of
any of these Hamiltonians, say $H_{-}$ then

\begin{equation}
\label{positiveSD1}
 \left<\alpha|H_{-}|\alpha\right>=\alpha\left<\alpha|\alpha\right>=\alpha
\end{equation}
where $\alpha$ is the eigenvalue of $|\alpha\left.\right>$. At the same time

\begin{equation}
\label{positiveSD2}
\left<\alpha|H_{-}|\alpha\right>=\left<\alpha|A^{\dagger}
A|\alpha\right>=\left|A|\alpha\left.\right>\right|^2
\end{equation}

where $\left|.\right|$ is the norm  of our Hilbert space, whose range is always
real.
Since the right hand side of \eref{positiveSD1} must be equal to the right hand side of
\eref{positiveSD2}, and by the positivity axiom of our metric space (Hilbert
space), $\left||x\left.\right>\right|\geq0$ for all the vectors
in our metric space \cite[pp.14]{Sakurai199309}, so it follows that $H_{-}$ is
positive semi-definite. It also follows that $H_-$ has eigenvalue zero
\textit{iff} $A$ annihilates a vector, which must be the ground state. The same
argument can be applied to $H_{+}$. For an alternative proof see \cite[Th. X25,
pp 180]{ReedSimon197510}.

It is straight forward to show that if $\psi^{-}$ is an eigenstate for $H_{-}$ with a positive eigenvalue $E^{-}$

\begin{equation}
 H_{-}\psi^{-}=A^{\dagger}A\psi^{-}=E^{-}\psi^{-}
\end{equation}

then $\psi^{+}=(E^{-})^{-1/2}(A\psi^{-})$ is a normalized eigenstate for $H_{+}$
with an eigenvalue $E^{+}=E^{-}$

\begin{equation}
 H_{+}\psi^{+}=\frac{1}{\sqrt{E^{-}}}AA^{\dagger}A\psi^{-}=E^{-}\psi^{+}
\end{equation}

 in the same manner $A^{\dagger}$ transforms an eigenstate of $H_{+}$ to an
eigenstate of $H_{-}$ with the same energy \cite{Sukumar1985l57,Arai199163}. This can be written as

\numparts

\begin{equation}
 Q\left(\begin{array}{c}
   \psi^{-}\\
0
  \end{array}\right)=
\left(\begin{array}{c}
 0\\
A\psi^{-}
 \end{array}\right)
\end{equation}

\begin{equation}
 Q^{\dagger}\left(\begin{array}{c}
0\\
\psi^{+}
  \end{array}\right)=
\left(\begin{array}{c}
A^{\dagger}\psi^{+}\\
0
 \end{array}\right)
\end{equation}

\endnumparts

 This result comes with no surprise for us, since we constructed our charge
operators \eref{charges} to obey our algebra \eref{algebra}, and we saw that
this algebra implies that the charges commute with the Hamiltonian
\eref{qhcommutator}, this commutativity is responsible for this degeneracy.

In literature, the linear operators $A$ and $A^{\dagger}$ are usually considered
in the following form

\numparts \label{shiftOperators}
 \begin{equation}
\label{A}
  A=\frac{d}{dx}+w(x)
 \end{equation}

\begin{equation}
\label{adjA}
 A^{\dagger}=-\frac{d}{dx}+w(x)
\end{equation}

\endnumparts

where $w(x)$ is an arbitrary function of $x$ and it is called the
\emph{superpotential}. Now, we can look at the differential form of our partner
Hamiltonians

\numparts \label{partnerH}
 \begin{equation}
H_{-}=A^{\dagger}A=-\frac{d}{dx^2}+w^2(x)-w'(x)
 \end{equation}
 \begin{equation}
H_{+}=AA^{\dagger}=-\frac{d}{dx^2}+w^2(x)+w'(x)
 \end{equation}

\endnumparts

the prime is $d/dx$, and we set $\hbar=2m=1$. The potentials in the above
Hamiltonians are called the \emph{supersymmetric partner potentials}
or superpartners, and they are given by

\begin{equation}
 V_{\pm}=w^2(x)\pm w'(x)
\end{equation}

this nonlinear differential equation is known as Ricatti equation. A node-less
solution to the Hamiltonians \eref{partnerH} can written in
terms of the superpotential

\begin{equation}
 \psi_0^{\pm}=N \exp\left(\pm\int^xw(t)dt\right)
\end{equation}

where $\psi_0^-$ is a solution of $H_{-}$ and $\psi_0^+$ is a solution of
$H_{+}$, the subscript is the number of nodes in the function, zero in this
case. $N$ is a normalization constant. By direct substitution we see that $A$
annihilates $\psi_0^{-}$, $A^{\dagger}$ annihilates $\psi_0^+$, consequently
they are solutions with eigenvalue zero

\numparts

\begin{equation}
  H_{-}\psi_0^-=A^{\dagger}(A\psi_0^-)=0
 \end{equation}

 \begin{equation}
  H_{+}\psi_0^+=A(A^{\dagger}\psi_0^+)=0
 \end{equation}

\endnumparts

but, these wave functions are not normalizable, i.e square integrable, at the
same time, actually they need not be normalizable at all,
but in the case when one of them is normalizable the supersymmetry is said to be
unbroken. If non of them is normalizable,
then non of the partner Hamiltonians has a zero eigenvalue and the supersymmetry
is broken \cite{Cooper1995267,Witten1982253,Cooper1983262}.

From now on we will consider the case where the supersymmetry is unbroken, i.e
one of $\psi_0^{-}$ and $\psi_0^{+}$ is normalizable, say $\psi_0^{-}$. In this
case $H_{-}$ has an eigenvalue zero. We saw earlier that if $H_{-}$ has an
eigenstate with a positive eigenvalue then $H_{+}$ also has the same eigenvalue
and vice versa. Since the lowest eigenvalue for $H_{-}$ is zero, then the lowest
eigenvalue of $H_{+}$ is the first positive eigenvalue of $H_{-}$. Moreover, the
operators $A$ and $A^{\dagger}$ connect the eigenstates of $H_{-}$ and $H_{+}$
except for the ground state of $H_{-}$
\cite{Sukumar1985l57,Arai199163,Cooper1995267}. This can be
summarized in the following equations

\numparts

\begin{equation}
 H_-\psi_n^-=E_n^-\psi_n^-\ \ \ \ \ \ \ n=0,1,2,...
\end{equation}

\begin{equation}
 H_+\psi_n^+=E_n^+\psi_n^+\ \ \ \ \ \ \ n=0,1,2,...
\end{equation}

\begin{equation}
E_{n+1}^-=E_n^+\ \ \ \ \ \ \ n=0,1,2,..., \ \ \ \ \ and\ \ \ \ E_0^-=0
\end{equation}

\begin{equation}
\psi_n^+=\frac{1}{\sqrt{E_{n+1}^{-}}}A\psi_{n+1}^-\ \ \ \ \ \ \ n=0,1,2,...
\end{equation}

\begin{equation}
\psi_{n+1}^-=\frac{1}{\sqrt{E_n^{+}}}A^{\dagger}\psi_{n}^+\ \ \ \ \ \ \
n=0,1,2,...
\end{equation}

\endnumparts

We notice here that the operators $A$ and $A^{\dagger}$ annihilate and create
nodes in wave functions, respectively, and this is a result of unbroken
supersymmetry. They are not ladder operators though, because ladder operators
move up and down the states of the same Hilbert space in
contrast to these operators which shift us between different Hilbert spaces, for this reason some authors call
them shift operators \cite{Montemayor1987}.

All the development so far depended on having a superpotential $w(x)$ which
generates all the previous results. But the unbroken supersymmetric structure
which has been considered so far can be constructed for any one dimensional
quantum system with at least one bound state, i.e
\newline\newline
\textbf{Every one-dimensional potential with bound states admits SUSY:}
\newline
Given a one dimensional quantum system with at least one bound state one can find a partner Hamiltonian
which has exactly the same discrete spectrum except for the ground state energy
of $H_-$ \cite{Sukumar19852917,Montemayor1989}.

Let us consider having a one dimensional potential $V_-$ which admits bound
state wave functions $\psi_n^-$, and the ground state energy is $E_0^-$, its
Hamiltonian $H_-$ given by

\begin{equation}
 H_-=-\frac{d^2}{dx^2}+V_-(x)-E_0^-
\end{equation}

can be factorized and be written as

\begin{equation}
 H_-=A^{\dagger}A
\end{equation}

where $A$ and $A^{\dagger}$ are defined as in \eref{A} and \eref{adjA},
respectively. The superpotential is

\begin{equation}
 w(x)=-\frac{d}{dx}\ln\psi_0^-
\end{equation}

where $\psi_0^-$ is the ground state wave function. The ground state energy of
$H_-$ is zero, and we can construct the supersymmetric partner Hamiltonian
$H_+$

\begin{equation}
 H_+=AA^{\dagger}
\end{equation}

All the results of (18) are also true for our constructed partner
Hamiltonians. This show us how to construct a
supersymmetric system from a one dimensional potential. In fact this technique
can be repeated again and again to obtain a sequence of Hamiltonians
where all the adjacent pairs of Hamiltonians are supersymmetric partners, the
cardinality of this sequence is the same as the dimensionality of the Hilbert
space of $H_{-}$, which is the number of bound states it possess
\cite{Sukumar1985l57,Sukumar19852917,Cao1991}.

\section{Shape Invariance}

We saw that supersymmetry allows us to construct and solve a hierarchy of
Hamiltonians
provided that we know the solution to one of the Hamiltonians in the sequence.
The question now is, Can supersymmetry help us solve a supersymmetric system?
The answer to this question was provided in 1983 by
Gendenshte\^in \cite{Gendenshtein1983}. Having supersymmetry is not a sufficient
condition for the system to be exactly solvable, because as we
saw we can construct a supersymmetric Hamiltonian for any potential with bound
state(s). Gendenshte\^in discovered another symmetry which if the
supersymmetric
system satisfies it will be an exactly solvable system, this symmetry is known
as shape invariance. If our potential satisfies shape invariance we can readily
write down its bound state spectrum, and with the help of the charge operators
we can find the bound state wave functions
\cite{Cooper1995267,Gendenshtein1983,Dutt1988}. It turned out that all the
potentials which were known to be exactly solvable until then have
the shape invariance symmetry.

If the supersymmetric partner potentials have the same dependence on $x$ but
differ in
a parameter, in such a way that they are related to each other by a change of
of that parameter, then they are said to be shape invariant. Gendenshte\^in
stated this
condition in this way,

\begin{equation}
\label{shapeInvariance}
 V_+(x,a_0)=V_-(x,a_1)+R(a_1)
\end{equation}

where $a_0$ is a parameter in our original potential whose ground state energy
is zero. $a_1=f(a_0)$ where $f$ is assumed to be an arbitrary function for the
time being, and $a_n=f^n(a_0)$ where $f^n$ is the composition of $f$ with itself
$n$ times. The remainder $R(a_1)$ can be dependent on the parametrization
variable $a_0$ but never on $x$. In this case $V_-$ is said to be shape
invariant, and we can readily find its spectrum, take a look at $H^n$ and
$H^{n+1}$,

\numparts
 \begin{equation}
  H^n=-\frac{d^2}{dx^2}+V_-(x,a_n)+\sum_{k=1}^nR(a_k)
 \end{equation}

\begin{eqnarray}
   H^{n+1}=-\frac{d^2}{dx^2}+V_-(x,a_{n+1})+\sum_{k=1}^{n+1}R(a_k) \\
= -\frac{d^2}{dx^2}+V_+(x,a_n)+\sum_{k=1}^nR(a_k)
\end{eqnarray}
\endnumparts

where in the last step we applied \eref{shapeInvariance}. Here we see that $H^n$
and $H^{n+1}$ are supersymmetric, so they have identical spectrum
except for the ground state energy of $H^n$ which is,

$$E_0^n=\sum_{k=1}^nR(a_k)$$

but from the supersymmetry arguments discussed in section \eref{SUSYsection} we
know that $H^n$ has the same spectrum of $H_-\equiv H^0$ except
for the first $n$ levels of $H_-$ which are missing from $H^n$, so the ground
state energy of the nth Hamiltonian $H^n$ is
equal to the nth energy $E_n^-$ of our original Hamiltonian $H_-$. So, the
spectrum of
$H_-$ is

\begin{equation}
 E_n^-=\sum_{k=1}^nR(a_k),\ \ \ \ \ \ \ \ \ \ E_0^-=0
\end{equation}

Supersymmetry together with shape invariance give us a formal expression for the
wave functions $\psi_n^-$ of $H_-$, for if we know the ground state
wave function $\psi_0^-(x,a_0)$ of $H_-(x,a_0)$, then the ground state wave
function of $H^n(x,a_n)$ is $\psi_0^-(x,a_n)$, but the operator
$A^{\dagger}(x,a_{n-1})$ move us from the Hilbert space of $H^n$ to the Hilbert
space of $H^{n-1}$, thus we can apply this operator repeatitively till
we reach the Hilbert space of $H_-=H^0$, we need to remember here that
$A^{\dagger}$ creates a node in the wave function each time we apply it, so
applying it $n$ times we will reach $\psi_n^-(x,a_0)$, hence

\begin{equation}
 \psi_n^-(x,a_0)=A^{\dagger}(x,a_{0})A^{\dagger}(x,a_{1})\ .\ .\ .\
A^{\dagger}(x,a_{n-2})A^{\dagger}(x,a_{n-1})\psi^-_0(x,a_n)
\end{equation}

where $A^{\dagger}(x,a_n)$ is defined as

\begin{equation}
 A^{\dagger}(x,a_n)=\frac{d}{dx}+w(x,a_n)=\frac{d}{dx}-\frac{d}{dx}\ln\psi_0^-(x
,a_n)
\end{equation}

This formal expression for the wave functions of shape invariant potentials were
first presented in \cite{Dutt1986295}, the explicit expressions
for the wave functions of all the well known shape invariant wave potentials
have been worked out in \cite{Dabrowska1988}. Supersymmetry together
with shape invariance can also be exploited to obtain the scattering matrices
\cite{Khare1988,Cooper1988}.

Different types of shape invariance have been studied in the literature, they
are classified depending on the relation among the parameters which creates the
shape invariant sequence of potentials. The first and the most dominant type is
known as translational shape invariance, where the potential parameter is
shifted, $a_1\equiv f(a_0)=a_0+\alpha$. This is the most
well studied type of shape invariance, and many of the generalizations of the
shape invariance concept have been worked specifically for this type. In fact it
turned out that all of the textbook examples of exactly solvable nonrelativistic
one dimensional
shape invariant quantum systems have this type of shape invariance.
Complete lists of translational shape invariant potentials together with their
wave functions and scattering matrices have been prepared
\cite{Cooper1995267,Dutt1988,Dabrowska1988,Khare1988}.

Translational shape invariance was the only known shape invariance type for a
long time, and some authors hypothesized that it is a necessary condition for
shape invariance \cite{Barclay1991}. In 1993, as new types of shape invariance
emerged, it was clear that translational shape invariance is not the only type.
The first of the new types to emerge is the scaling shape invariance
\cite{Khare1993,Barclay1993}, $a_1=qa_0$, $0<q<1$. There are also two nonlinear
transformations which have been introduced together with the previous one in
\cite{Barclay1993}, $a_1=qa_0^p$, $p \in \mathbb{Z}$, $0<q<1$, and
$a_1=qa_0/(1+pa_0)$, $0<q$,$p<1$, $pa_0\ll1$. The eigenvalues, eigenfunction
functions and transmission coefficients have been obtained algebraically for
potentials with such shape invariance \cite{Khare1993,Barclay1993}.

Cyclic shape invariance has been introduced in the following years
\cite{Gangopadhyaya1996,Sukhatme1997}. In this type the parameters repeat
themselves after cycle of p elements, $a_p=a_0$.

In all the types, except for the translational shape invariance, the potentials
are not obtained in closed form, i.e in terms of elementary functions, they are
obtained in series form.

Another important generalization of the ordinary shape invariance relation
\eref{shapeInvariance}  is to have shape invariance in multi-steps.

Shape invariance has also been generalized in another direction. In 1987, in an
attempt to classify shape invariant potentials, Cooper et al.\cite{Cooper1987}
introduced the idea of translational shape invariance in an ``n'' arbitrary but
finite number of parameters, they were not successful at finding a solution for
such a system, they were even pessimistic about the existence of a solution. It
was an open problem for more than a decade, later on, by skillfully exploiting
some properties of the Ricatti equation a solution was shown to exist and worked
out\cite{Carinena20003467}.

Many successful efforts have been put to show the underlying algebraic structure
of the shape invariance symmetry, the associated Lie algebras have also been
identified \cite{Balantekin1998}, and in the case of the non translational shape
invariance nonlinear generalizations of Lie algebras have been obtained
\cite{Balantekin19982,Chaturvedi1998109}. 

The technique of factorizing the Hamiltonian and using shape invariance to solve
the schr\"{o}dinger equation is not at all new, Schr\"{o}dinger himself used this
technique to solve the hydrogen atom, Dirac also used it to solve the harmonic
oscillator. Of course the ideas of supersymmetry and the supergroups were not
known back then.

Later on, the technique of Schr\"{o}dinger and Dirac was rejuvenated by Infeld
and Hull and later they summarized their work in their infamous review
\cite{InfeldHull} were they gave the name for the method, ``The Factorization
Method''. They summarized the procedure of this technique to solve second order
differential equations, they studied its range of applicability and classified
the factorization types into six types (transformation between the types exist),
they also prepared a table for these six possible factorizations which can be
used to solve the differential equations simply by identifying to which
factorization type it
belongs \cite{InfeldHull}. For the historical progress of this method see
\cite{Mielnik2004}.
\newline \newline
\textbf{Every Infeld and Hull factorizable potential is shape invariant but the converse is not true: }
\newline
As far as solving differential equations is concerned, the factorization
technique offers almost the same power as the supersymmetry with
\textit{translational} shape invariance approach. The supersymmetry and shape
invariance offer physical insight and more group theoretic ideas of why such
systems are exactly solvable. In fact shape invariance, in general, offers more
than the factorization method, since the factorization method treats only the
translational shape invariance. The equivalence of the factorization method to
supersymmetry with \textit{translational} shape invariance is easy to see, equation (3.1.2) in \cite{InfeldHull} is exactly equivalent to \eref{shapeInvariance} with translation of the shape invariance parameter \cite{Montemayor1989,Carinena2000}.

The table prepared by Infeld and Hull, though complete for most purposes, is not
the most general table possible, as they did not consider the most general
solution of Ricatti equations. The most general solution is worked out in
\cite{Carinena2000}.
\newline\newline
\textbf{Shape invariance is a sufficient but not a necessary condition for exact solubility:}
\newline
Gendenshte\^in suggested that all exactly solvable potentials must be shape
invariant \cite{Gendenshtein1983}, but now the relation among exact
solubility, shape invariance and supersymmetry is much clearer. One counter example is enough to refute Gendenshte\^in suggestions, and many such examples have been constructed \cite{Cao1991,Cooper1987,cooper1989}, the most important of these counter examples is Natanzon class of potentials which are, in general, not shape invariant \cite{Cooper1987,cooper1989}.

\begin{figure}[htp]
\centering
    \includegraphics[width=.5 \textwidth]{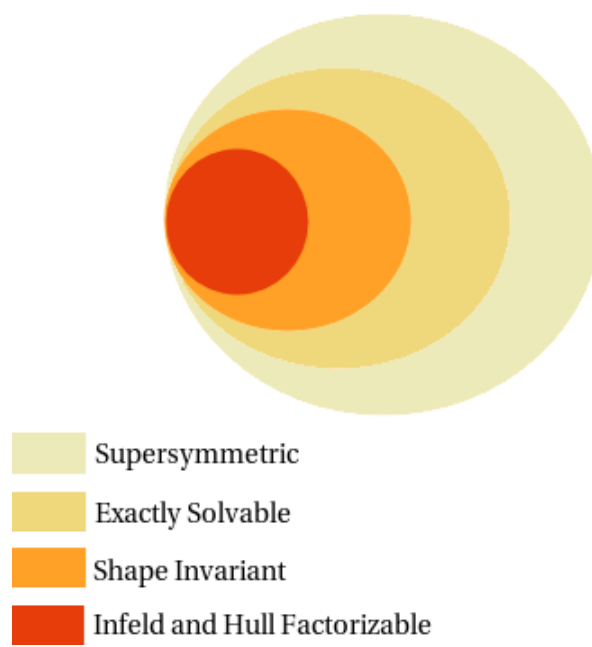}
  \caption{A Venn diagram showing the relation among, supersymmetric, exactly
solvable, shape invariant, and Infeld and Hull factorizable potentials}
\label{venn}
\end{figure}

\section{Conclusion}
In this letter we defined, supersymmetric quantum mechanics, and shape invarinace. We also showed the relation which connects them to exact solubility and the factorization method. Fig.\ref{venn} summarizes these relations.

\ack
We would like to thank the ARO for financial support. One of us (Mustafa) would like to thank Professor M.Shikakhwa for the enlightening discussions and comments. 

\section*{References}
\bibliographystyle{unsrt}
\bibliography{letter}
\end{document}